\documentclass[10pt,letterpaper]{article}

\usepackage{amsmath,amsfonts,bm}

\def\eqref#1{equation~\ref{#1}}

\def\1{\bm{1}}

\def\ve{{\bm{e}}}

\def\vh{{\bm{h}}}

\DeclareMathAlphabet{\mathsfit}{\encodingdefault}{\sfdefault}{m}{sl}
\SetMathAlphabet{\mathsfit}{bold}{\encodingdefault}{\sfdefault}{bx}{n}

\newcommand{\R}{\mathbb{R}}

\usepackage{arxiv}
\usepackage{graphicx}
\usepackage{amssymb}
\usepackage{amsmath}
\usepackage{bm}
\usepackage{color}
\usepackage{cite}
\usepackage{colortbl}
\usepackage{longtable}
\usepackage{subfigure}
\usepackage{multirow,comment}
\usepackage{hyperref}
\usepackage{wrapfig}
\usepackage{algorithmic}
\usepackage[ruled,vlined]{algorithm2e}
\usepackage{enumitem} 
\usepackage[applemac]{inputenc}
\usepackage[T1]{fontenc}

\def\etal{\emph{et al.}}
\definecolor{citecolor}{RGB}{34,139,34}

\newcommand{\be}{\begin{equation}}
\newcommand{\ee}{\end{equation}}
\newcommand{\bea}{\begin{eqnarray}}
\newcommand{\eea}{\end{eqnarray}}

\makeatletter\renewcommand\paragraph{\@startsection{paragraph}{4}{\z@}
  {.5em \@plus1ex \@minus.2ex}{-.5em}{\normalfont\normalsize\bfseries}}\makeatother

\newlength\savewidth\newcommand\shline{\noalign{\global\savewidth\arrayrulewidth
  \global\arrayrulewidth 1pt}\hline\noalign{\global\arrayrulewidth\savewidth}}

\makeatletter\renewcommand\paragraph{\@startsection{paragraph}{4}{\z@}
  {.5em \@plus1ex \@minus.2ex}{-.5em}{\normalfont\normalsize\bfseries}}\makeatother

\author{{\hspace{1mm}Junfeng Wu}, {\hspace{1mm}Dawei Leng}, {\hspace{2mm}Lurong Pan*}\\
 \emph{AIDD Group}\\
 Global Health Drug Discovery Institute, Beijing, China \\
 \texttt{\{junfeng.wu, lurong.pan\}@ghddi.org}\\}

\title{ParaVS: A Simple, Fast, Efficient and Flexible Graph Neural Network Framework for Structure-Based Virtual Screening}

\newcommand{\bullets}[1]{\hspace*{1pc}\textbf{\emph{-- #1}}}

\begin{document}

\maketitle

\begin{abstract}
Structure-based virtual screening (SBVS) is a promising \emph{in silico} technique that integrates computational methods into drug design. An extensively used method in SBVS is molecular docking. However, the docking process can hardly be computationally efficient and accurate simultaneously because classic mechanics scoring function is used to approximate, but hardly reach, the quantum mechanics precision in this method. In order to reduce the computational cost of the protein-ligand scoring process and use data driven approach to boost the scoring function accuracy, we introduce a docking-based SBVS method and, furthermore, a deep learning non-docking-based method that is able to avoid the computational cost of the docking process. Then, we try to integrate these two methods into an easy-to-use framework, ParaVS, that provides both choices for researchers. Graph neural network (GNN) is employed in ParaVS, and we explained how our in-house GNN, HagNet, works and how to model ligands and molecular targets. To verify our approaches, cross validation experiments are done on two datasets, an open dataset Directory of Useful Decoys: Enhanced (DUD.E) and an in-house proprietary dataset without computational generated artificial decoys (NoDecoy). On DUD.E we achieved a state-of-the-art AUC of $0.981$ and a state-of-the-art enrichment factor at $2\%$ of 36.2; on NoDecoy we achieved an AUC of $0.974$. We further finish inference of an open database, Enamine REAL Database (RDB), that comprises over 1.36 billion molecules in 4050 core-hours using our ParaVS non-docking method (ParaVS-ND). The inference speed of ParaVS-ND is about $3.6e5\ molecule\ /\ core-hour$, while this number of a conventional docking-based method is around 20, which is about 16000 times faster. The experiments indicate that ParaVS is accurate and computationally efficient and can be generalized to different molecular targets for virtual screening. Moreover, ParaVS is simple and flexible since most AI components can be substituted or updated as per specific demand and future development. ParaVS-ND has been released as a free online inference service in \url{http://aidd.ghddi.org/sbvs/} and code will be released soon.
\end{abstract}

\vspace{-.5em}
\section{Introduction}

\begin{figure}[t]
\centering
\includegraphics[height=25em]{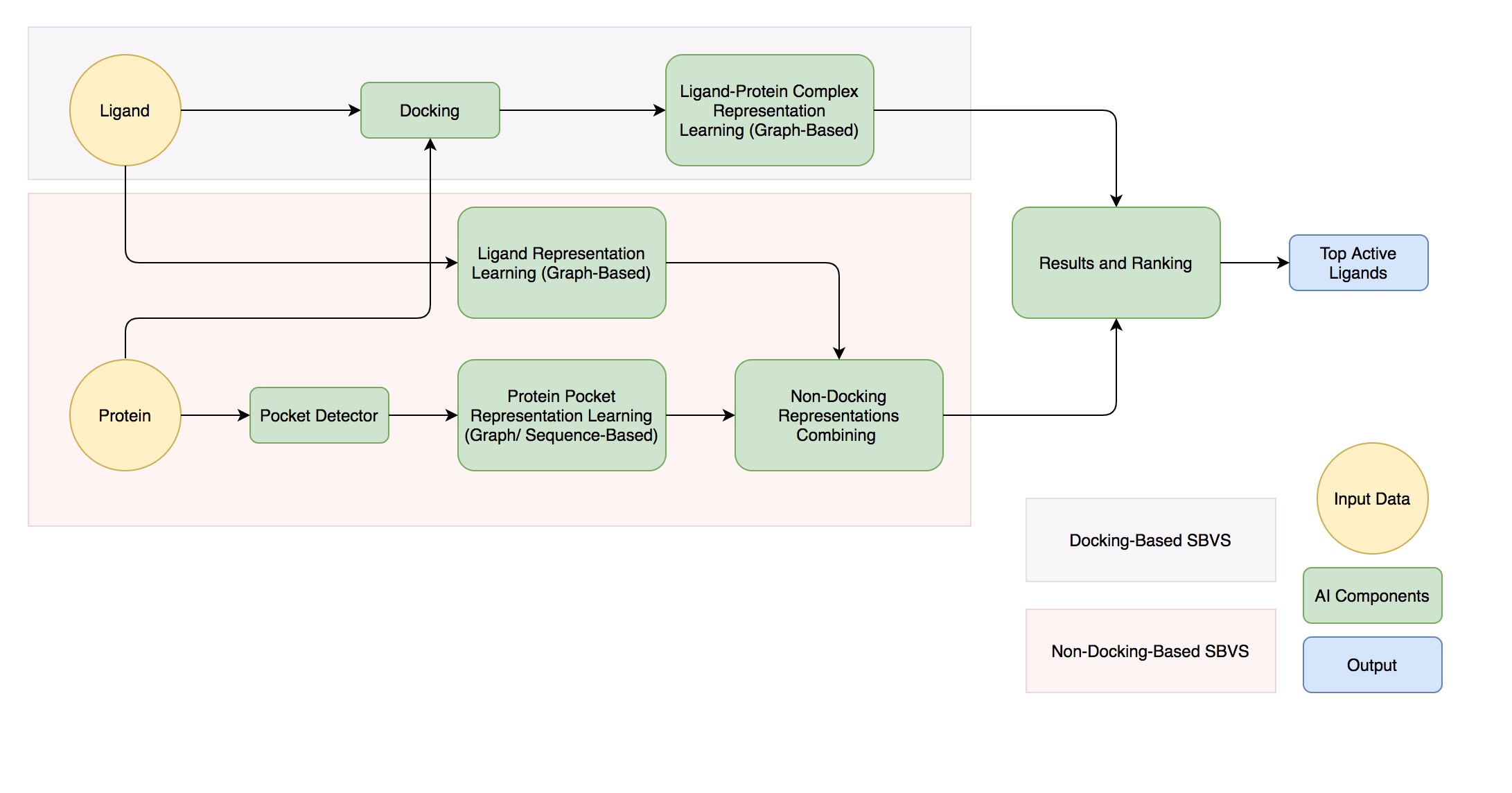}
\caption{Structure-Based Virtual Screening Framework}
\vspace{-1em}
\label{fig:sbvs_framework}
\end{figure}

Drug discovery process is a well known time-consuming and expensive task. The discovery of new drugs goes through several development stages, each of which could give out thousands of millions of molecules for screening\cite{Hecht2009}. Formerly, random screening and empirical advices were utilized for this task. Then, high-throughput screening (HTS) improves drug discovery process by offering automated and much quicker screening of large chemical libraries against a molecular target\cite{Iram2015}. However, HTS is still considered to be time-consuming, laborious, expensive with a relatively low screening accuracy\cite{Cheng2012, Schneider2010}.

With the rapid development of computer science, computational methods are extensively used in drug design now. Specifically, virtual screening (VS) is one of the most promising \emph{in silico} techniques that works as a filter, and structure-based virtual screening (SBVS) is a robust and powerful VS that predicts the interaction between a compound and a molecular target protein\cite{Ghosh2006, Bissantz2001}. SBVS utilizes the three-dimensional (3D) structure information of molecular target, hence, it normally produces a better performance than ligand-based VS, which does not contain any structure information\cite{Arciniega2014}.

A commonly used methodology in SBVS is molecular docking\cite{Ghosh2006, Drwal2013, Kitchen2004}. It is a procedure that inserts a ligand into a particular region, as known as a pocket, of a molecular target protein. Most docking programs accomplish this task by performing pocket searching, ligand insertion and pose determination using molecular mechanics (MM) and molecular dynamics (MD) and other heuristic algorithms\cite{VENKATACHALAM2003289}. Then, multiple possible results are output to ranking algorithms to determine a final docked target protein structure\cite{Kitchen2004, Durrant2010,Kinnings2011}, namely a protein-ligand complex, thus, biological and chemical properties can be inferred for screening. 

Furthermore, deep learning (DL) becomes a prominent tool in VS. Comparing to traditional machine learning (ML) algorithms\cite{Hecht2009, Cheng2012}, such as neural network\cite{Durrant2010}, support vector machine (SVM)\cite{Kinnings2011} and random forest (RF)\cite{Ballester2010}, deep learning models require minimum feature engineering. Ideally speaking, if enough data is provided, representations or features can be learned directly from dataset without any human design\cite{Bengio2009, Bengio2013, LeCun2015}. An early successful DL example is Quantitative Structure-Activity Relationship (QSAR) \cite{Vilar2008}, which operates over molecular fingerprints. Wallach \etal \cite{Wallach2015} find a way to apply 3D convolutional neural network (3D-CNN) to a protein-ligand complex which is represented on a 3D grid. Ragoza \etal \cite{Ragoza_2017} further extended it to include active and inactive binding poses classification. These 3D-CNN methods successfully outperform previous works and also improve the accuracy of predicting absolute binding affinities\cite{Wang2002, Ballester2010, Durrant2011}.

Nevertheless, docking-based VS is not a perfect answer. The main problems of docking are:
most docking processes are time and computing power consuming; docking results are not highly reliable; balanced and high-quality data is limited. 
In order to resolve them, non-docking methods that avoid using any docking programs are proposed. A common framework is to integrate representations of ligands and proteins into a single neural network without knowing 3D binding structures. Karimi \etal \cite{Karimi2018} introduced their DeepAffnity to predict the pIC50 of protein-ligand complexes with sequences of proteins and  simplified molecular-input line-entry system(SMILES) \cite{Weininger1988} strings of ligands as inputs. Gao \etal \cite{Gao2018} developed a siamese network consists of a recurrent neural network (RNN) and a graph neural network (GNN), and they also brought attention mechanism\cite{vaswani2017attention} into their model to mitigate the problem of generalizability, which is a common problem in SBVS that the performance drops notably if the testing ligand and protein are not seen in the training set. Lee \etal \cite{Lee2019} also reported their models working on sequences of proteins and SMILES strings of ligands. Nguyen \etal \cite{Nguyen2019} and Lim \etal \cite{Lim2019} further developed GNN based models for SBVS tasks.

In this paper, we propose a docking-based (ParaVS-Dock) and a non-docking-based (ParaVS-ND) method for SBVS tasks, and establish a framework containing both of them, as illustrated in Figure~\ref{fig:sbvs_framework}. We evaluate both methods on two large datasets, an open dataset Directory of Useful Decoys: Enhanced (DUD.E)\cite{Mysinger2012} and a Global Health Drug Discovery Institute (GHDDI) in-house dataset NoDecoy. We also evaluate our methods on DUD, a subset of DUD.E. On DUD.E, we achieved an AUC$_{ParaVS-Dock}=0.926$ and a state-of-the-art AUC$_{ParaVS-ND} = 0.981$. On NoDecoy dataset, our results are AUC$_{ParaVS-Dock}=0.911$ and AUC$_{ParaVS-ND} = 0.974$, proving the robustness and generalizability of our methods. We did not do any deep learning fine-tuning on all our experiments, which implies a potential of getting better performances. Besides, in order to demonstrate ParaVS computationally efficient, we perform inference on a large database, Enamine REAL Database(RDB)~\cite{RDB2007}, and successfully boost the inference speed more than 16000 times by circumventing the docking process.

We summarize our contributions as follows:
\vspace{-.4em}
\begin{enumerate}[label=(\roman*)]
\item We proposed a SBVS framework, ParaVS, with a docking-based (ParaVS-Dock) and a non-docking based (ParaVS-ND) method. ParaVS is highly optimizable as the AI components, shown as green blocks in Figure~\ref{fig:sbvs_framework}, can easily be switched or updated.
\item We evaluated our methods on two datasets, an open dataset and an in-house proprietary dataset, and both acquired AUC$_\text{mean}> 0.9$, demonstrating their capability and generalizability.
\item We investigated how to model a molecular target, and explained why non-docking-based methods are generally better than docking-based ones.
\item Our methods are of low computational cost. Table~\ref{tab:docking_time_cost} lists the time we used for our experiments. By circumventing the docking process, ParaVS-ND can be surprisingly faster than docking-based methods, especially the inference speed has boosted from 20 sample per CPU core-hour to $3.6e5$.
\end{enumerate}
\vspace{-.4em}

This paper is briefly organized as: Section~\ref{sec:method} explains GNN and our models in detail; Section~\ref{sec:settings} describes the datasets we used, and how we perform our experiments; Section~\ref{sec:results} presents the results of our methods and comparisons to methods listed in other literatures; Section~\ref{sec:conclusion} summarizes our conclusions and also discusses several general problems of SBVS.
\begin{table}[t]
  \centering
  \small
  \begin{tabular}{c|c| c | c}
    \multicolumn{2}{c|}{Description}
    & ParaVS-Dock & ParaVS-ND \\
    \shline
    \multirow{4}{*} {time (core-hour)} & \emph{DUD.E} docking & $7e4$ & -\\
    & \emph{DUD.E} LOO CV (10 epochs, GPU) & $200$ & $1600$\\
    & \emph{RDB} docking & $6.7e7$ & - \\
    & \emph{RDB} inference & $1500$ & $4050$\\
    \hline
    \multirow{1}{*} {speed (sample / core-hour)} & inference speed & $20$ & $3.6e5$ \\ 
  \end{tabular}
  \caption{RDB\cite{RDB2007} contains $1.36e9$ compounds and DUD.E contains $1.4e6$ compounds. The docking process is done by \emph{Autodock Vina1.1.2}\cite{Trott2009}, and all listed tasks are run by parallel processing clusters consist of 12 \emph{Intel(R) Xeon(R) Gold 5118 CPU @ 2.30GHz} CPU, except that the DUD.E leave-one-out(LOO) cross-validation(CV) experiment is done by GPU for 10 epochs. We use core-hour to measure computational cost, which refers to the number of processing units (cores) used to run a job multiplied by the duration in hours.}
  \label{tab:docking_time_cost}
\end{table}

\section{Methods}\label{sec:method}

Our goal is to develop both a docking-based and a non-docking-based structure-based virtual screening(VS) method. Given this goal, a brief introduction to general GNN model and our in-house GNN model is provided. Then, our SBVS methods are discussed comprehensively.

\subsection{Graph Neural Network}
Graph neural network(GNN) is widely used in drug discovery process now, as ligands and proteins can be modeled as graphs by nature \cite{zhou2019graph}. A graph is a pair $G=(V,E)$, where $V$ is a set of nodes and $E$ is a set of edges, and an edge is a set of paired nodes. GNNs model ligands and proteins as graphs, in which the nodes are atoms and the edges are defined simply by connecting atoms that lie within a predefined cutoff distance $c$, then GNNs represent each atom via an atom embedding layer $\vh_u \in \R^H$ and each edge via an edge embedding layer $\ve_{uv} \in \R^H$.

Most GNNs in this area can be summarized as a \textbf{two-stage architecture}, as known as the message-passing framework \cite{Gilmer2017}: 1. Propagate node information among each other by neighborhood aggregation on each layer. 2. Form the whole graph representation by a read-out function. Each layer of such GNNs can be written as:
\begin{eqnarray}
\vec h_v^{(l+1)} & = & f_\theta^{(l)}(\vec h_v^{(l)}, \{\vec x_e:e\in \mathcal{N}_E(v)\}, \{\vec h_{v'}^{(l)}:v'\in \mathcal{N}_V(v)\}).
\label{eq:gnn}
\end{eqnarray}
where $\vec h_v^{(l)}$ is the node feature vector of node $v$ at layer $l$. $\mathcal{N}_V(v)$ and $\mathcal{N}_E(v)$ denote sets of nodes and edges connecting to the node $v$, and $f_\theta^{(l)}$ is a parameterized function.
The readout function $R$ pools node features from the final iteration $K$ to obtain the entire graph's representation $h_G$:
\begin{eqnarray}
\vec h_G & = & R(\{h_v^K | v \in G\}).
\label{eq:gnn}
\end{eqnarray}

In our implementation, we use our in-house developed GNN, HagNet\cite{dawei2021}. HagNet can be formally described as:
\begin{eqnarray}
\vec h_v^{(l+1)} & = & \phi (\textbf{concat}(h_v^{l}, (\sum_{u \in \mathcal{N} (v)}h_u^l + max_{u \in \mathcal{N}(v)} h_u^l)))
\label{eq:gnn}
\end{eqnarray}
where $\phi$ is a multilayer perceptron(MLP) network and \textbf{concat} layer concatenates features.
\begin{figure}[t]
\centering
\includegraphics[height=22em]{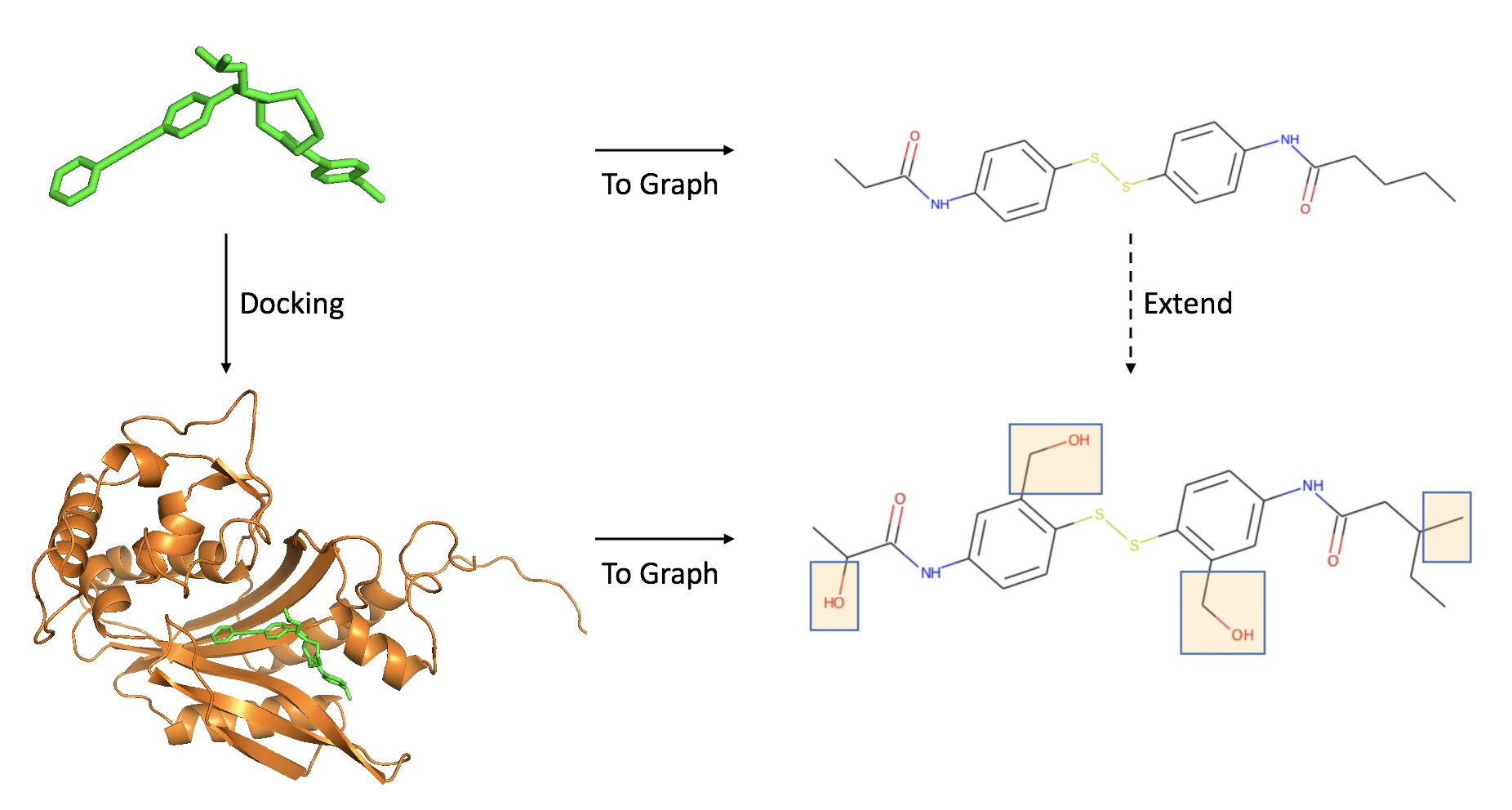}
\caption{Docked protein-ligand complex graph extension procedure}
\vspace{-1em}
\label{fig:dock_extend_procedure}
\end{figure}

\begin{wrapfigure}{R}{0.48\textwidth}
	\begin{minipage}{0.48\textwidth}
		\begin{algorithm}[H]
			\caption{ParaVS-Dock. $f$ is the graph neural network we need to train. $dist(i,j)$ is the distance between node $i$ and $j$. $d_j$ is the minimum distance of node $j$ to the ligand, defined as $d_j:=\min_{i \in V_c,j}dist(i,j)$}
			\label{alg:db_alg}
			\begin{algorithmic}
                \STATE {\bfseries input:} ligand atoms $V_c$, protein atoms from a docked protein-ligand complex $V_p$, cutoff distance $c$
                \STATE $V := \{i | i \in V_c\} \cup \{j | j \in V_p, d_j < c\}$
                \STATE $E := \{\{i, j\} | i, j \in V, dist(i, j) < c, i \neq j \}$
                \STATE $G := (V, E)$
                \STATE $y \gets f(G)$
				\STATE  {\bfseries return }$y$
			\end{algorithmic}
		\end{algorithm}
  \end{minipage}
  \vspace{-1em}
\end{wrapfigure}
\subsection{ParaVS-Dock}
ParaVS-Dock is a docking-based method that takes a docking result generated by \emph{Autodock Vina1.1.2\cite{Trott2009}} as an input, and output the whole docked protein-ligand complex representation via a single GNN. The pseudo-codes are presented in Algorithm~\ref{alg:db_alg}. First, using the output of docking programs, which is the 3D structure of a protein-ligand complex, to form a graph by setting atoms as nodes, and connecting each two nodes if they are within a cutoff distance $c$. Instead of modeling the whole protein-ligand complex into a graph, we solely put the atoms of the ligand into a graph and ignoring atoms of the protein for now. Then, the graph is extended by adding atoms of the protein whose distances to any atom of the ligand are within the cutoff distance $c$, as illustrated in Fig~\ref{fig:dock_extend_procedure} where atoms of the protein are shown with orange background color. Finally, a GNN is applied to the extended graph. The idea behind is that the interaction between the protein and the ligand is vital and informative for the final prediction.

We did not model the whole protein-ligand complex as a graph for the following reasons:
\vspace{-.4em}
\begin{enumerate}[label=(\roman*)]
\item A protein size is normally much larger than a ligand size. A typical protein contains over 2000 atoms, while the number of atoms in a ligand is about 50. Even a protein pocket, which is significantly smaller than a protein, can easily contain more than 500 atoms, still ten times the size of a ligand. Therefore, a protein-ligand complex graph will be dominated by the protein, increasing the difficulty for GNNs to differentiate active and inactive compounds. 
\item On the contrary to (i), the number of proteins is much smaller than the number of compounds in most SBVS datasets, in which over 1k or even 10k compounds paired with a single protein, making most protein-ligand complex graphs in the training set structurally similar to each other and further increases the difficulty of training.
\item A docked protein-ligand complex graph can be too large for a GNN to train, causing serious efficiency and computing power problems.
\end{enumerate}
\vspace{-.4em}

\subsection{Non-Docking-Based}
\begin{figure}[t]
\centering
\includegraphics[width=46em]{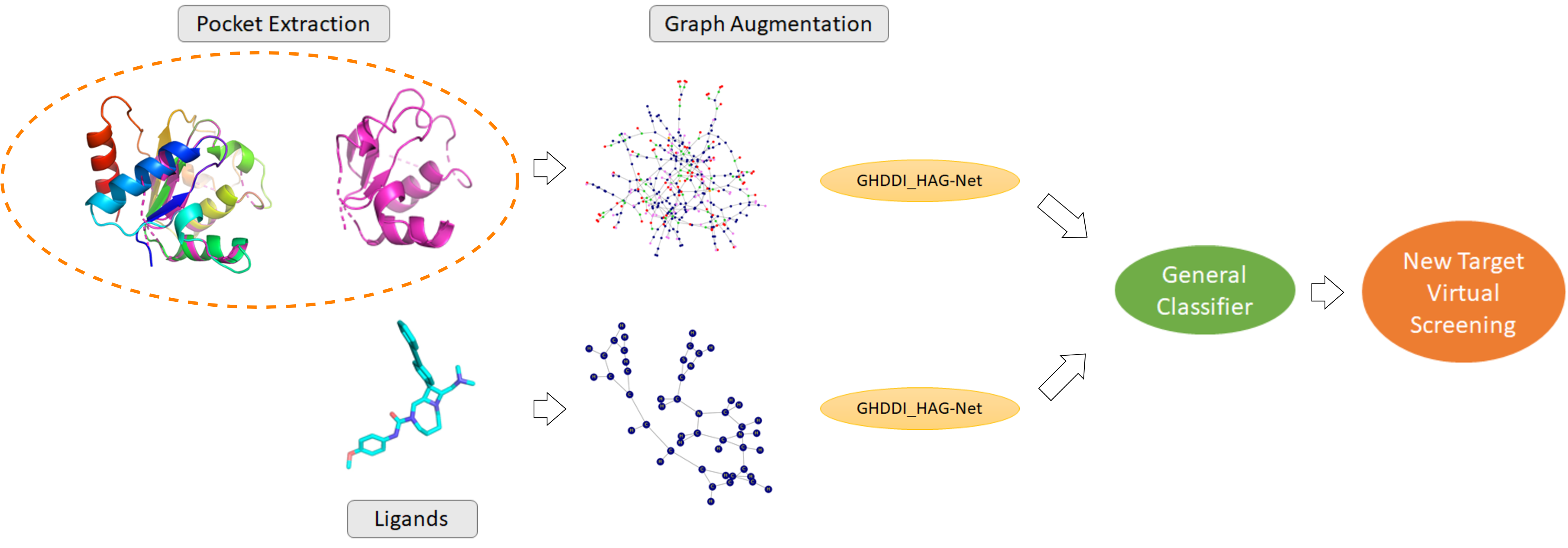}
\caption{ParaVS-ND: Non-docking-based SBVS procedure}
\vspace{-1em}
\label{fig:nd_procedure}
\end{figure}

Docking-based methods generally suffer from three main problems:
1. Most docking processes are time and computing power consuming while researchers often have to process a large amount of compounds. As listed in Table~\ref{tab:docking_time_cost}, it costs us months to finish DUD.E docking process with multiple parallel computing clusters. It is either too time-consuming or less approachable for those who are not accessible to high-performance computing.
2. Relying on heuristic algorithms, empirical functions, machine learning, or other techniques, docking programs are not giving out analytical solutions. Errors are introduced during the docking stage, then propagate through follow-up models, and eventually harm the final performances.
3. It is fairly difficult to acquire high-quality data with binding poses.
Thus, non-docking-based methods can be a vital supplement to SBVS tasks.

\begin{wrapfigure}{R}{0.48\textwidth}
	\begin{minipage}{0.48\textwidth}
		\begin{algorithm}[H]
			\caption{ParaVS-ND. $f_c$ is the GNN we need to train for ligand and $f_p$ is the GNN for pocket, $\vh_u$ is the atom embedding layer}
			\label{alg:nd_alg}
			\begin{algorithmic}
                \STATE {\bfseries input:} ligand graph $G_c = (V_c, E_c)$, protein pocket graph $G_p = (V_p, E_p)$; cutoff distance $c$
                \STATE $h_p^u \gets \vh_u(V_p)$
                \STATE $h_c^u \gets \vh_u(V_c)$
                \STATE $h_p \gets f_p(h_p^u, E_p)$
                \STATE $h_c \gets f_c(h_c^u, E_c)$
                \STATE $h \gets concat(h_p, h_c)$
                \STATE $y \gets MLP(h)$
				\STATE  {\bfseries return }$y$
			\end{algorithmic}
		\end{algorithm}
	\end{minipage}
\end{wrapfigure}

\begin{figure}[t]
\centering
\includegraphics[width=48em]{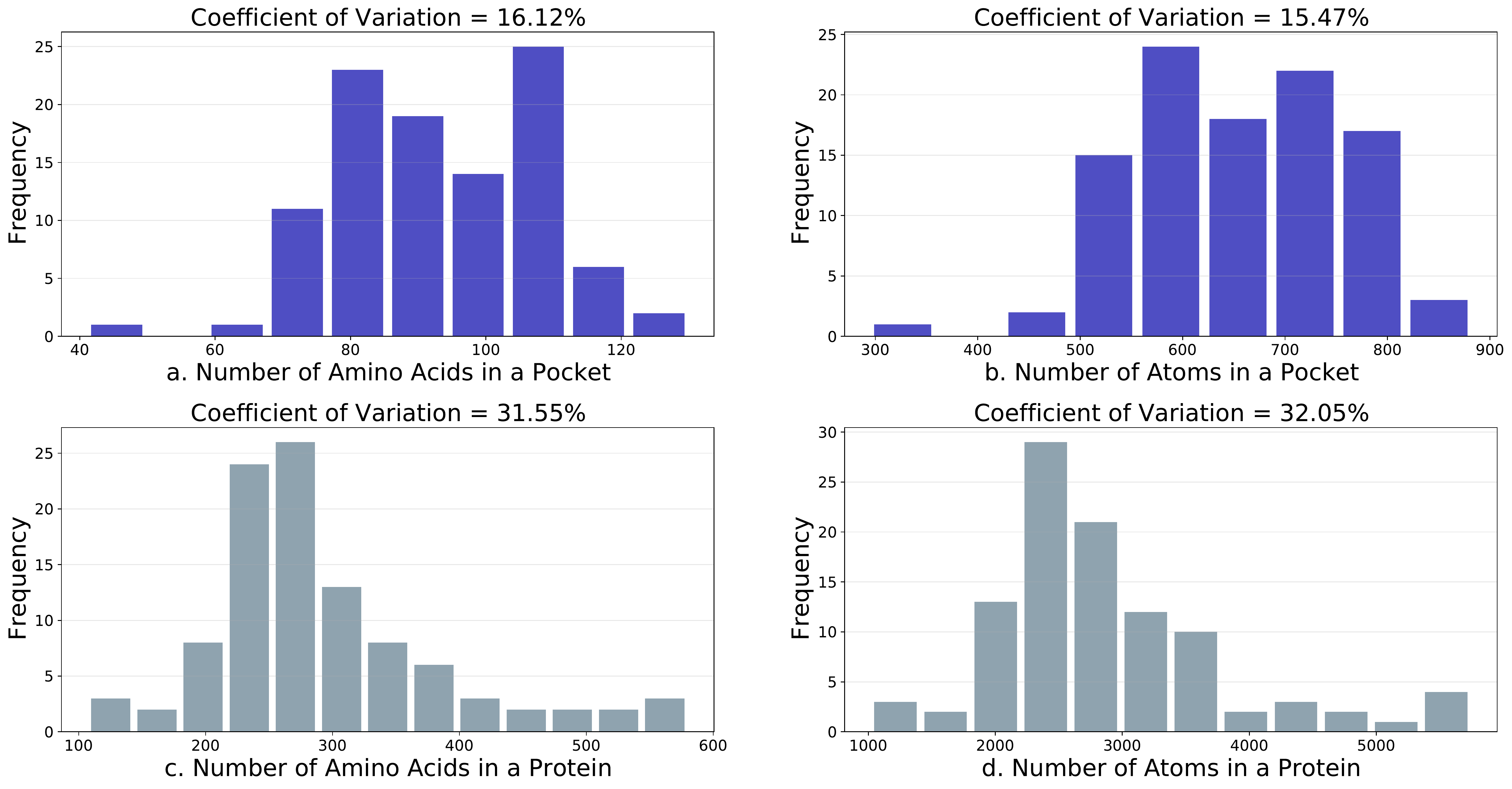}
\caption{(\textbf{a}) and (\textbf{b}) denote the distributions of numbers of amino acids and atoms in a pocket; (\textbf{c}) and (\textbf{d}) denote the distributions of numbers of amino acids and atoms in a protein. Distributions in (\textbf{c}) and (\textbf{d}) are right-skewed with higher variation and higher CoV comparing to (\textbf{a}) and (\textbf{b}), thus, they are harder to be modeled by GNN.}
\vspace{1em}
\label{fig:coef_var}
\end{figure}

However, contradicting to ligand modeling, how to model a protein remains a main problem that is unclear and worth discussing in SBVS tasks. Figure~\ref{fig:coef_var} shows the statistical analysis of this problem on our in-house NoDecoy dataset, which is described in Table~\ref{tab:dataset_stats} and Section~\ref{sec:settings} in detail.

Coefficient of Variation (CoV) in Figure~\ref{fig:coef_var} is defined as the ratio of the standard deviation $\sigma$ to the mean $\mu$: $c_v = \sigma / \mu$. It is a standardized measure of dispersion of a probability distribution, a smaller $c_v$ indicates that the data concentrates more on a certain range and easier to be modeled.

\bullets{Amino Acids or Atoms?} The first decision happens between whether we should model in amino acids or atoms. A protein is made up of amino acids, and an amino acid is made up of atoms. Therefore, a protein can be modeled in both. We select to use atoms for the reason that a typical SBVS dataset, including DUD.E and our in-house NoDecoy, contains much more proteins than compounds, as reported in Table~\ref{tab:dataset_stats}. Modeling in amino acids has a serious issue of lacking enough training data for the amino acid embedding layer. On the contrary, this problem can be naturally resolved by sharing the atom embedding layer with the ligand part, since ligands are consisted of atoms but \emph{NOT} amino acids.

\bullets{Protein or Pocket?} Figure~\ref{fig:coef_var}.a and Figure~\ref{fig:coef_var}.b report $c_v$ about $16\%$ and Figure~\ref{fig:coef_var}.c and Figure~\ref{fig:coef_var}.d report $c_v$ about $32\%$. From Figure~\ref{fig:coef_var}, It is clear that, whether we choose to model in amino acids or atoms, to model pocket only instead of a full protein is easier for GNNs. Additionally, Figure~\ref{fig:coef_var} also indicates that a protein is about 4 to 6 times larger than a pocket, whether in amino acids or atoms, resulting in more time and computing power consuming.

The pseudo-codes of ParaVS-ND is presented in Algorithm~\ref{alg:nd_alg}. Firstly, two individual graphs are formed via the same methodology stated in the docking-based method using the ligand and the protein separately. They are initiated by the same atom embedding layer, then two individual GNNs are applied to get the representations. Finally, the representations are concatenated as an input to a classifier neural network, which is a MLP in our implementation. Note that weights are not shared between the two GNNs(except the atom embedding layer), since ligands and proteins are different from each other in attributes, sizes and features. The framework is also displayed in Figure~\ref{fig:nd_procedure}. While we seek to maintain it as simple as possible, multiple optional modules are available, such as replacing the MLP with a CNN or a RNN module, adding attention mechanisms to the MLP, etc.

\section{Experimental Settings}\label{sec:settings}
\paragraph{Dataset}
\begin{table}[t]
\centering
\small
\begin{tabular}{ccccc}
Number of & \# of Proteins & \# of Compounds & \# of Active Ligands & Active Ratio \\
\shline
\multirow{1}{*}{DUD.E} & 102 & 1434015 & 22805 & 1.59\% \\
\hline
\multirow{1}{*}{DUD} & 38 & 631474 & 10491 & 1.66\% \\
\hline
\multirow{1}{*}{NoDecoy} & 185 & 284792 & 222687 & 78.19\% \\
\end{tabular}
\vspace{.5em}
\caption{DUD.E and NoDecoy dataset descriptions, Active Ratio $= \frac{\#\ of\ Active\ Ligands}{\#\ of\ Compounds}$}
\label{tab:dataset_stats}
\end{table}

We analyze our models and methods on two datasets, a public dataset and a dataset combining both public and in-house proprietary data. With such setting, we are trying to evaluate our approach's generalizability and reproducibility. The number of proteins and ligands of both datasets are listed in Table~\ref{tab:dataset_stats}. Additionally, we perform inference on RDB with both ParaVS-Dock and ParaVS-ND to compare the computational efficiency, as reported in Table~\ref{tab:docking_time_cost}.

\bullets{DUD.E}: An enhanced and rebuilt version of DUD(Directory of Useful Decoys)\cite{Mysinger2012}. It contains 102 protein targets. We train and evaluate our model on both full DUD.E dataset and its subset DUD, which contains 38 proteins out of 102 as shown in Table~\ref{tab:dataset_stats}.

\bullets{NoDecoy}: A combination of DUD.E, excluding its computer generated decoys, and our in-house proprietary dataset which contains biochemical data of 261987 compounds and 83 targets using $IC_{50}=10 \mu M$ as cutoff to differentiate active and inactive compounds. We exclude decoys from DUD.E in order to eliminate implicit computer generated patterns, and to balance the proportion of positive and negative samples. Then we combine the remaining compounds with our proprietary dataset, which contains no decoys by nature.

\bullets{RDB}: Enamine REAL database (RDB)\cite{RDB2007} is the largest enumerated database which comprises over 1.36 billion synthetically feasible molecules. We use RDB for inference only in this paper to simulate a regular screening task, since it is such a large database that can reveal the computational efficiency of ParaVS-ND.

\paragraph{Cross Validation}
The performances of ParaVS are assessed in two types of cross validation(CV).

\bullets{Leave-One-Out(LOO) CV}: Leave-one-out cross validation is a technique for assessing how the method will generalize to independent data. Specifically, the model will leave one protein and all compounds pairing with it out in rotation, and train on all other remaining proteins and compounds, then do model testing on the left-out protein and compounds. We compare our results with DeepVS\cite{Pereira2016} and Lim \etal \cite{Lim2019}.

\bullets{K-fold CV}: In $k$-fold CV, the dataset is randomly partitioned into $k$ equal sized subsets. Of the $k$ subsets, a single subset is retained as the validation set, and other $k-1$ subsets are used as the training set. The CV process is repeated $k$ times, with each subset used as the validation set exactly once. In our experiments, $k=5$.

\paragraph{Evaluation Metrics}
To evaluate our approaches, we select two extensively used metrics for comparison: the enrichment factor (EF) and the area under the Receiver Operating Characteristics curve (AUC).

\bullets{AUC}: AUC is also written as AUROC in some literatures, and is one of the most important evaluation metrics for classification problems\cite{Fawcett2006}.

\bullets{EF}: Since the tested compounds are ranked by score, EF is used to indicate how good the prediction is on the top $x\%$ ranked compounds. The EF at $x\%$ is computed as:

\begin{eqnarray}
EF_{x\%} = \frac{actives\ at\ x\%}{compounds\ at\ x\%} \times \frac{total\ compounds}{total\ actives}
\label{eq:ef}
\end{eqnarray}

\paragraph{Deep Learning Settings}
We try to maintain the training and testing procedure as simple as possible, avoiding too many tricks in order to verify the robustness and generalizability of our methods.

\bullets{Pre-training}: All models are trained from scratch without any pre-training.

\bullets{Fine-tuning}: GNN hyper-parameters are set to be the same for all the experiments, and we did not do grid search or any other fine-tuning on our models. Better performances are expected to be achievable with more deliberate works.

\bullets{Optimizer}: SGD with a constant learning rate.

\bullets{Computation Power}: All experiments are individually run on a single P100 GPU. The computation cost of our method is relatively low.

  
\section{Results and Discussion}\label{sec:results}
\subsection{DUD and DUD.E LOO experiment}

\begin{table}[t]
\centering
\small
\begin{tabular}{c|ccc|cccc}
  \textbf{} & EF$_\text{max}^{DUD.E}$ & EF$_\text{2\%}^{DUD.E}$ & EF$_\text{20\%}^{DUD.E}$ & EF$_\text{max}^{DUD}$ & EF$_\text{2\%}^{DUD}$ & EF$_\text{20\%}^{DUD}$\\
\shline
\multirow{1}{*}{ParaVS-Dock} & 34.4 & 22.4 & 4.4 & 31.3 & 19.2 & 4.3 \\
\multirow{1}{*}{ParaVS-ND} & 62.6 & \textbf{39.7} & 4.9 & \textbf{60.1} & \textbf{36.2} & \textbf{4.8} \\
\hline
\multirow{1}{*}{DeepVS-ADV\cite{Pereira2016}} & - & - & - & 16.0 & 6.6 & 3.1 \\
\multirow{1}{*}{Lim \etal \cite{Lim2019}} & - & 33.5 & - & - & - & - \\
\multirow{1}{*}{Lim \etal \cite{Lim2019} w/o attention} & - & 30.3 & - & - & - & - \\
\multirow{1}{*}{Ragoza \etal \cite{Ragoza_2017}} & - & 19.4 & - & - & - & - \\
\multirow{1}{*}{Torng \etal \cite{Torng2019}} & - & 19.4 & - & - & - & - \\
\end{tabular}  \\
\vspace{.5em}
\caption{Mean Enrichment Factor(EF) comparison. The presented EF value is a \emph{mean} value computed by averaging over proteins in LOO CV experiments. All EF$^{DUD.E}$ values are performed on DUD.E and EF$^{DUD}$ values are on DUD.}
\label{tab:ef2}
\end{table}

\begin{table}[t]
\centering
\small
\begin{tabular}{cc|ccccc}
& \textbf{AUC$_{mean}$} & DUD.E & DUD \\
\shline
\multirow{2}{*}{} & ParaVS-Dock & 0.926 & 0.911 \\
& ParaVS-ND & \textbf{0.981} & 0.974 \\
\hline
\multirow{3}{*}{} & DeepVS-ADV\cite{Pereira2016} & - & 0.81 \\
& Lim \etal \cite{Lim2019} & 0.968 & - \\
& Lim \etal \cite{Lim2019} w/o attention & 0.936 & - \\
& AtomNet \cite{Wallach2015} & 0.855 & - \\
& Ragoza \etal \cite{Ragoza_2017} & 0.868 & - \\
& Torng \etal \cite{Torng2019} & 0.886 & - \\
& Gonczarek \etal \cite{Gonczarek_2018} & 0.904 & - \\
\end{tabular}  \\
\vspace{.5em}
\caption{Mean AUC comparison. The mean value is computed by averaging over proteins in LOO CV experiments. It shows that ParaVS-ND outperforms methods in other literatures, and an AUC$_\text{mean}=0.981$ is state-of-the-art as far as we know.}
\label{tab:auc_comparison}
\end{table}

\begin{figure}[t]
\centering
\includegraphics[width=46em]{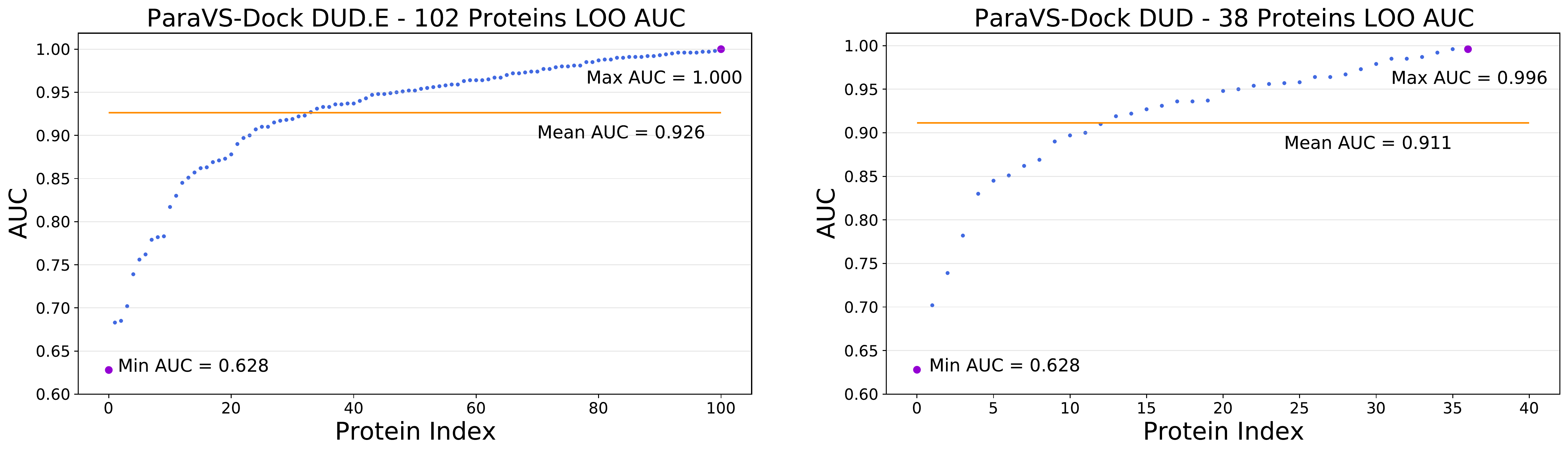}
\caption{ParaVS-Dock leave-one-out(LOO) cross-validation(CV) performance, each point represent an AUC score from LOO CV, and proteins are sorted in ascending order by AUC. \textbf{Left}: DUD.E: AUC$_\text{max}=1.000$, AUC$_\text{min}=0.628$, AUC$_\text{mean}=0.926$. \textbf{Right}: DUD: AUC$_\text{max}=0.996$, AUC$_\text{min}=0.628$, AUC$_\text{mean}=0.911$.}
\label{fig:loo_auc_dude}
\vspace{-1em}
\end{figure}

The results of ParaVS-Dock are reported in Figure~\ref{fig:loo_auc_dude}. Figure~\ref{fig:loo_auc_dude}(left) shows the LOO CV AUC of each of 102 target proteins from DUD.E, most of which achieves an AUC $>0.90$, and only 10 AUCs are under $0.80$. Figure~\ref{fig:loo_auc_dude} (right) shows each of 38 target proteins from DUD.

\begin{figure}[t]
\vspace{1em}
\centering
\includegraphics[width=46em]{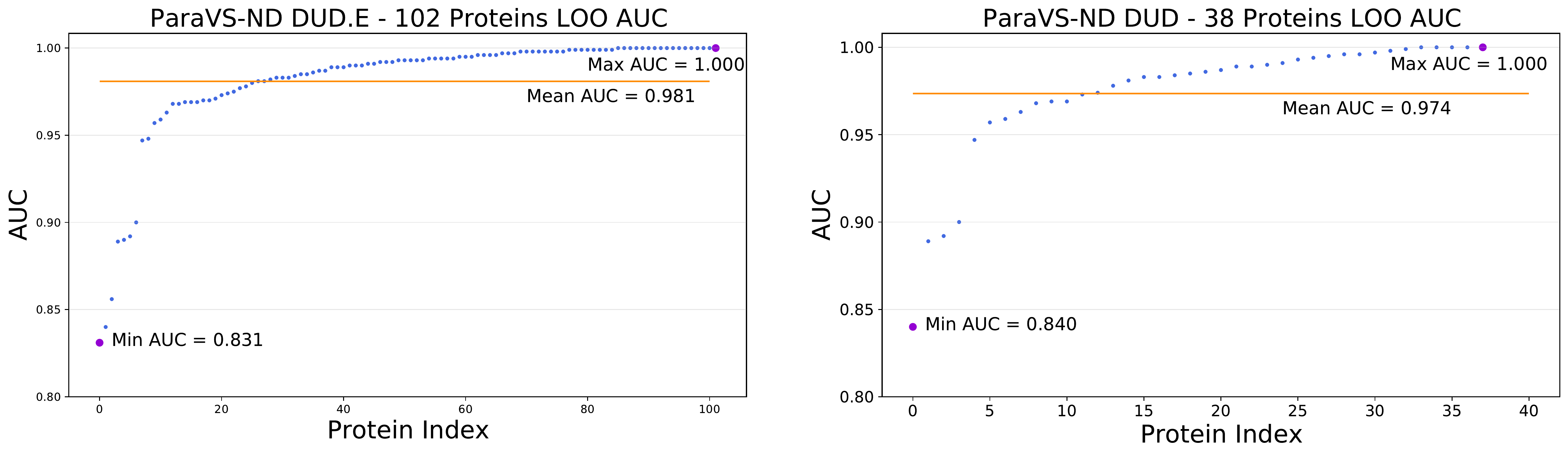}
\caption{ParaVS-ND leave-one-out(LOO) cross-validation(CV) performance, each point represent an AUC score from LOO CV, and proteins are sorted in ascending order by AUC. \textbf{Left}: DUD.E: AUC$_\text{max}=1.000$, AUC$_\text{min}=0.831$, AUC$_\text{mean}=0.981$. \textbf{Right}: DUD: AUC$_\text{max}=1.000$, AUC$_\text{min}=0.840$, AUC$_\text{mean}=0.974$.}
\label{fig:loo_auc_dude_nd}
\vspace{-1em}
\end{figure}

Next, we test ParaVS-ND with the same settings, as shown in Figure~\ref{fig:loo_auc_dude_nd}. Comparing to ParaVS-Dock, AUC$_\text{mean}$ has boosted from $0.926$ to $0.981$, and most AUCs are above $0.95$ with AUC$_\text{min}$ improved from $0.628$ to $0.831$, revealing that ParaVS-ND is more robust and of greater generalizability.

Comparisons of AUC and EF to methods listed in other literatures are reported in Table~\ref{tab:ef2} and Table~\ref{tab:auc_comparison}. They show that both ParaVS-Dock and ParaVS-ND perform well. Specifically, ParaVS-ND acquires a mean EF$_\text{2\%} = 39.7$ on DUD.E, mean EF$_\text{2\%} = 36.2$ on DUD, and an AUC$_\text{mean}=0.981$ on DUD.E and AUC$_\text{mean} = 0.974$ on DUD, all are state-of-the-art as far as we know at the moment.

\subsection{NoDecoy 5-Fold experiment} 
\begin{figure}[t]
\centering
\includegraphics[width=46em]{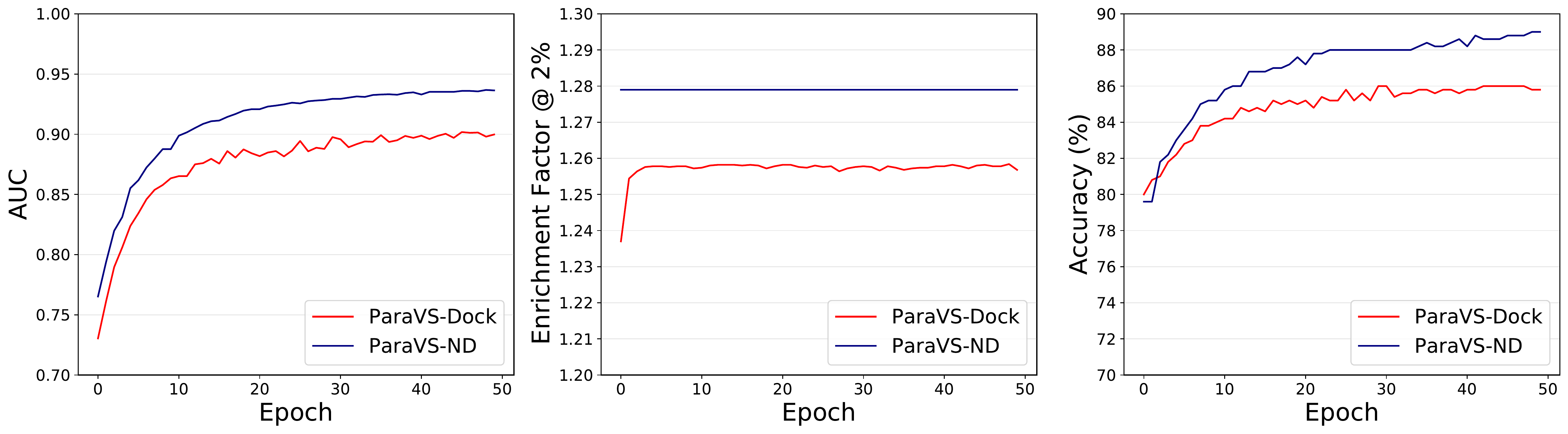}
\caption{ParaVS-Dock {\emph{vs.}} ParaVS-ND performance in the $k-$fold cross validation, where $k=5$ in our experiment. Metrics are averaged on each epoch by $k$ folds. \textbf{Left}: AUC curve. \textbf{Middle}: Enrichment factor @ 2\% curve. \textbf{Right}: Accuracy curve.}
\label{fig:kfold_d_vs_nd}
\vspace{-1em}
\end{figure}

To evaluate our methods and framework at production stage, we eliminate decoys from DUD.E and integrate its true active ligands into our in-house dataset to form a new dataset, namely NoDecoy. Afterwards, we perform a 5-fold CV experiment by a random split. As illustrated in Figure~\ref{fig:kfold_d_vs_nd}, we acquired an AUC$_\text{mean}=0.902$ for ParaVS-Dock and AUC$_\text{mean}=0.937$ for ParaVS-ND. Considering the dataset contains no decoys and active ratio $=78.19\%$, which is much more balanced than DUD.E whose active ratio $=1.59\%$ as listed in Table~\ref{tab:dataset_stats}, we find the performance quite satisfying. EF$_\text{2\%}$ maintains a constant number $1.279$ because it is the maximum value we can achieve, since in Equation~\ref{eq:ef} we have:

\begin{eqnarray}
\frac{total\ ligands}{total\ actives} = \frac{1}{Active\ Ratio} = \frac{1}{78.19\%} = 1.279
\label{eq:why_ef_max}
\end{eqnarray}

In addition to AUC and EF, we also evaluate the prediction accuracy. Figure~\ref{fig:kfold_d_vs_nd} reveals that: ParaVS-ND outperforms ParaVS-Dock in all metrics we assessed; performances of ParaVS-ND raise and converge faster; the AUC curve of the ParaVS-ND fluctuates less, indicating a better model stability; the AUC and accuracy curves of both methods are still raising, it is expected that better performances may be reached given more training epochs.

\subsection{RDB inference speed}
Table~\ref{tab:docking_time_cost} reports that ParaVS-Dock is faster than ParaVS-ND by itself, but the docking process is too time-consuming that ParaVS-Dock spent most of its running time on docking and becomes incomparable to ParaVS-ND efficiency-wise. The RDB test is valuable for the reason that performing inference on thousands of millions of compounds is not a rare thing in drug discovery process. The core-hours ParaVS-ND took for RDB is a fairly good result that about $3.6e5$ compounds can be screened in an hour with a single CPU, and the speed is scalable with parallel processing clusters.

\section{Conclusion and Discussion}\label{sec:conclusion}
We summarize our main conclusion as follows:

- \emph{Non-docking-based methods possess advantages over docking-based ones based on results from our experiments.}

- \emph{GNN can play a significant role in drug design, and an increasing number of methods start to use GNN to model molecules now.}

- \emph{Datasets from virtual screening process can be fairly large, thus, it is vital for VS methods to balance performance and efficiency.}

We further provide answers to a few important questions people are interested in:

\emph{\textbf{Is GNN a solution to SBVS tasks?}} Yes, although GNN would \emph{NOT} be the only answer and there are definitely other models worth investigating. Since ligands and target proteins can be modeled as graphs in a straightforward way, and theories of GNN are developing rapidly, we expect SBVS tasks to benefit more from GNN models in the future.

\emph{\textbf{Is the non-docking-based method better than the docking-based method in all aspects?}} No, there are a few problems worth noticing:

\bullets{Model size}: The non-docking-based model is almost twice the size of the docking-based one since it contains two independent GNNs. Although it can be imagined that reducing the model size by trimming models skillfully can be a help, the docking-based method is preferred in a model size sensitive task.

\bullets{Computing power cost and speed}: There is a strong motivation to reduce computing time and save computing power in drug discovery process, while the non-docking-based method takes more time in both training and inference. But things change when the full procedure taken into consideration, as the docking process, known as a time-consuming step, can be circumvented in the non-docking-based method. In short, the docking-based method is faster \emph{ONLY} when docked protein-ligand complexes are provided.

\bullets{Pocket Detection}: We acquired pockets from proteins along with the dataset. However, this is \emph{NOT} always the case. If pockets are not provided, we have to implement pocket detection, which can be done by various methods\cite{Guilloux2009, Stepniewska2019, Cui2019}. The situation is similar to docking that the analytical solution is not available and, although much more efficient and faster than docking, pocket detection takes time.

Drug discovery has long been a tough task. It has progressed significantly ever since computational methods got into the field. Hopefully, our work can inspire more thoughts about SBVS, and the analysis and discussions about ligand and protein modeling can slightly bridge the gap between drug discovery and computer science.

{\small
\bibliographystyle{ieee}
\bibliography{reference}

\begin{thebibliography}{10}\itemsep=-1pt

\bibitem{Arciniega2014}
M.~Arciniega and O.~Lange.
\newblock Improvement of virtual screening results by docking data feature
  analysis.
\newblock {\em Journal of chemical information and modeling}, 54, 05 2014.

\bibitem{Ballester2010}
P.~Ballester and J.~Mitchell.
\newblock A machine learning approach to predicting protein-ligand binding
  affinity with applications to molecular docking.
\newblock {\em Bioinformatics (Oxford, England)}, 26:1169--75, 03 2010.

\bibitem{Bengio2009}
Y.~Bengio.
\newblock Learning deep architectures for ai.
\newblock {\em Foundations}, 2:1--55, 01 2009.

\bibitem{Bengio2013}
Y.~Bengio, A.~Courville, and P.~Vincent.
\newblock Representation learning: A review and new perspectives.
\newblock {\em IEEE transactions on pattern analysis and machine intelligence},
  35:1798--1828, 08 2013.

\bibitem{Bissantz2001}
C.~Bissantz, G.~Folkers, and D.~Rognan.
\newblock Protein-based virtual screening of chemical databases. 1. evaluation
  of different docking/scoring combinations.
\newblock {\em Journal of medicinal chemistry}, 43:4759--67, 01 2001.

\bibitem{Cheng2012}
T.~Cheng, Q.~Li, Z.~Zhou, Y.~Wang, and S.~Bryant.
\newblock Structure-based virtual screening for drug discovery: a
  problem-centric review.
\newblock {\em The AAPS journal}, 14:133--41, 03 2012.

\bibitem{Cui2019}
Y.~Cui, Q.~Dong, D.~Hong, and X.~Wang.
\newblock Predicting protein-ligand binding residues with deep convolutional
  neural networks.
\newblock {\em BMC Bioinformatics}, 20, 02 2019.

\bibitem{Drwal2013}
M.~Drwal and R.~Griffith.
\newblock Combination of ligand- and structure-based methods in virtual
  screening.
\newblock {\em Drug Discovery Today: Technologies}, 10:395, 06 2013.

\bibitem{Durrant2010}
J.~Durrant and J.~McCammon.
\newblock Nnscore: A neural-network-based scoring function for the
  characterization of protein-ligand complexes.
\newblock {\em Journal of chemical information and modeling}, 50:1865--71, 10
  2010.

\bibitem{Durrant2011}
J.~Durrant and J.~McCammon.
\newblock Nnscore 2.0: A neural-network receptor-ligand scoring function.
\newblock {\em Journal of chemical information and modeling}, 51:2897--903, 11
  2011.

\bibitem{Fawcett2006}
T.~Fawcett.
\newblock Introduction to roc analysis.
\newblock {\em Pattern Recognition Letters}, 27:861--874, 06 2006.

\bibitem{Gao2018}
K.~Gao, A.~Fokoue, H.~Luo, A.~Iyengar, and P.~Zhang.
\newblock Interpretable drug target prediction using deep neural
  representation.
\newblock pages 3371--3377, 07 2018.

\bibitem{Ghosh2006}
S.~Ghosh, A.~Nie, J.~An, and Z.~Huang.
\newblock Structure-based virtual screening of chemical libraries for drug
  discovery.
\newblock {\em Current opinion in chemical biology}, 10:194--202, 07 2006.

\bibitem{Gilmer2017}
J.~Gilmer, S.~Schoenholz, P.~Riley, O.~Vinyals, and G.~Dahl.
\newblock Neural message passing for quantum chemistry.
\newblock 04 2017.

\bibitem{Gonczarek_2018}
A.~Gonczarek, J.~M. Tomczak, S.~Zaręba, J.~Kaczmar, P.~Dąbrowski, and M.~J.
  Walczak.
\newblock Interaction prediction in structure-based virtual screening using
  deep learning.
\newblock {\em Computers in Biology and Medicine}, 100:253–258, Sep 2018.

\bibitem{Hecht2009}
D.~Hecht and G.~Fogel.
\newblock Computational intelligence methods for docking scores.
\newblock {\em Current Computer - Aided Drug Design}, 5, 03 2009.

\bibitem{Iram2015}
W.~Iram and t.~anjum.
\newblock {\em Production of Cyclosporine A by Submerged Fermentation}, pages 1
  -- 28.
\newblock 11 2015.

\bibitem{Karimi2018}
M.~Karimi, D.~Wu, Z.~Wang, and Y.~Shen.
\newblock Deepaffinity: Interpretable deep learning of compound protein
  affinity through unified recurrent and convolutional neural networks, 06
  2018.

\bibitem{Kinnings2011}
S.~Kinnings, N.~Liu, P.~Tonge, R.~Jackson, L.~Xie, and P.~Bourne.
\newblock A machine learning-based method to improve docking scoring functions
  and its application to drug repurposing.
\newblock {\em Journal of chemical information and modeling}, 51:408--19, 02
  2011.

\bibitem{Kitchen2004}
D.~Kitchen, H.~Decornez, J.~Furr, and J.~Bajorath.
\newblock Docking and scoring in virtual screening for drug discovery: Methods
  and applications.
\newblock {\em Nature reviews. Drug discovery}, 3:935--49, 12 2004.

\bibitem{Guilloux2009}
V.~Le~Guilloux, P.~Schmidtke, and P.~Tuffery.
\newblock Fpocket: An open source platform for ligand pocket detection.
\newblock {\em BMC bioinformatics}, 10:168, 02 2009.

\bibitem{LeCun2015}
Y.~LeCun, Y.~Bengio, and G.~Hinton.
\newblock Deep learning.
\newblock {\em Nature}, 521:436--44, 05 2015.

\bibitem{Lee2019}
I.~Lee, J.~Keum, and H.~Nam.
\newblock Deepconv-dti: Prediction of drug-target interactions via deep
  learning with convolution on protein sequences.
\newblock {\em PLOS Computational Biology}, 15:e1007129, 06 2019.

\bibitem{dawei2021}
D.~Leng.
\newblock Heterogeneous aggregation graph network for molecule property
  prediction., preprint 2021.

\bibitem{Lim2019}
J.~Lim, S.~Ryu, K.~Park, Y.~Choe, J.~Ham, and W.~Kim.
\newblock Predicting drug-target interaction using a novel graph neural network
  with 3d structure-embedded graph representation.
\newblock {\em Journal of Chemical Information and Modeling}, 59, 08 2019.

\bibitem{Mysinger2012}
M.~Mysinger, M.~Carchia, J.~Irwin, and B.~Shoichet.
\newblock Directory of useful decoys, enhanced (dud-e): Better ligands and
  decoys for better benchmarking.
\newblock {\em Journal of medicinal chemistry}, 55:6582--94, 06 2012.

\bibitem{Nguyen2019}
T.~Nguyen, H.~le, and S.~Venkatesh.
\newblock Graphdta: prediction of drug-target binding affinity using graph
  convolutional networks, 06 2019.

\bibitem{Pereira2016}
J.~C. Pereira, E.~R. Caffarena, and C.~N. dos Santos.
\newblock Boosting docking-based virtual screening with deep learning.
\newblock {\em Journal of Chemical Information and Modeling},
  56(12):2495--2506, 2016.
\newblock PMID: 28024405.

\bibitem{Ragoza_2017}
M.~Ragoza, J.~Hochuli, E.~Idrobo, J.~Sunseri, and D.~R. Koes.
\newblock Protein–ligand scoring with convolutional neural networks.
\newblock {\em Journal of Chemical Information and Modeling}, 57(4):942–957,
  Apr 2017.

\bibitem{Schneider2010}
G.~Schneider.
\newblock Virtual screening: An endless staircase?
\newblock {\em Nature reviews. Drug discovery}, 9:273--6, 04 2010.

\bibitem{RDB2007}
A.~Shivanyuk, S.~Ryabukhin, A.~Bogolyubsky, D.~Mykytenko, A.~Chuprina,
  W.~Heilman, A.~Kostyuk, and A.~Tolmachev.
\newblock Enamine real database: Making chemical diversity real.
\newblock {\em Chimica Oggi}, 25:58--59, 11 2007.

\bibitem{Stepniewska2019}
M.~Stepniewska-Dziubinska, P.~Zielenkiewicz, and P.~Siedlecki.
\newblock Detection of protein-ligand binding sites with 3d segmentation, 04
  2019.

\bibitem{Torng2019}
W.~Torng and R.~Altman.
\newblock Graph convolutional neural networks for predicting drug-target
  interactions.
\newblock {\em Journal of Chemical Information and Modeling}, 2019, 10 2019.

\bibitem{Trott2009}
O.~Trott and A.~Olson.
\newblock Software news and update autodock vina: Improving the speed and
  accuracy of docking with a new scoring function, efficient optimization, and
  multithreading.
\newblock {\em Journal of computational chemistry}, 31:455--61, 11 2009.

\bibitem{vaswani2017attention}
A.~Vaswani, N.~Shazeer, N.~Parmar, J.~Uszkoreit, L.~Jones, A.~N. Gomez,
  L.~Kaiser, and I.~Polosukhin.
\newblock Attention is all you need, 2017.

\bibitem{VENKATACHALAM2003289}
C.~Venkatachalam, X.~Jiang, T.~Oldfield, and M.~Waldman.
\newblock Ligandfit: a novel method for the shape-directed rapid docking of
  ligands to protein active sites.
\newblock {\em Journal of Molecular Graphics and Modelling}, 21(4):289 -- 307,
  2003.

\bibitem{Vilar2008}
S.~Vilar, G.~Cozza, and S.~Moro.
\newblock Medicinal chemistry and the molecular operating environment (moe):
  application of qsar and molecular docking to drug discovery.
\newblock {\em Current topics in medicinal chemistry}, 8(18):1555—1572, 2008.

\bibitem{Wallach2015}
I.~Wallach, M.~Dzamba, and A.~Heifets.
\newblock Atomnet: A deep convolutional neural network for bioactivity
  prediction in structure-based drug discovery.
\newblock 10 2015.

\bibitem{Wang2002}
R.~Wang, L.~Lai, and S.~Wang.
\newblock Further development and validation of empirical scoring functions for
  structure-based binding affinity prediction.
\newblock {\em Journal of computer-aided molecular design}, 16:11--26, 02 2002.

\bibitem{Weininger1988}
D.~Weininger.
\newblock Smiles, a chemical language and information system. 1. introduction
  to methodology and encoding rules.
\newblock {\em Journal of Chemical Information and Computer Sciences},
  28:31--36, 02 1988.

\bibitem{zhou2019graph}
J.~Zhou, G.~Cui, Z.~Zhang, C.~Yang, Z.~Liu, L.~Wang, C.~Li, and M.~Sun.
\newblock Graph neural networks: A review of methods and applications, 2019.

\end{thebibliography}
}

\end{document}